\documentclass[twocolumn,showpacs,amsmath,nofootinbib,pra,aps,amssymb,longbibliography,superscriptaddress]{revtex4-2} 
\usepackage[T1]{fontenc}
\usepackage[utf8]{inputenc}
\usepackage{times}
\usepackage{color} 
\usepackage{array}
\usepackage{amssymb,amsmath}
\usepackage{amsbsy}
\usepackage[pdftex]{graphicx} 
\usepackage{bm} 
\usepackage{float}
\usepackage{dcolumn}
\usepackage{booktabs}

\usepackage[unicode,breaklinks]{hyperref}
\hypersetup{
    unicode=true,
    plainpages=false, 
    colorlinks=true,
    linkcolor=blue,
    citecolor=blue,
    filecolor=black,
    urlcolor=blue
}
\usepackage{url}
\usepackage{verbatim}

\newcommand{\add}{a_{dd}}

\newcommand{\br}{\mathbf{r}}
\newcommand{\bq}{\mathbf{q}}

\newcommand{\bx}{\mathbf{x}}

\newcommand\gammaQF{\gamma_\mathrm{QF}}

\begin{document}
 
\title{Acoustic fluctuations of two-dimensional dipolar supersolids} 
 	\author{P.~B.~Blakie} 
\affiliation{%
	Dodd-Walls Centre for Photonic and Quantum Technologies, Dunedin 9054, New Zealand}
\affiliation{Department of Physics, University of Otago, Dunedin 9016, New Zealand}
\date{\today} 
\begin{abstract}  
We investigate the microscopic structure of the long-wavelength acoustic excitations of two-dimensional dipolar supersolids with triangular and honeycomb crystal order. Using Bogoliubov--de Gennes calculations, we determine the density, phase, current, and lattice-displacement fluctuations associated with the three acoustic modes. We develop a procedure to extract the lattice displacement directly from the microscopic density perturbation, allowing the modes to be identified as one transverse and two longitudinal branches. The displacement field further separates the density fluctuations into contributions from crystal strain and particle transport relative to the lattice. This reveals that the lower longitudinal mode can have large strain and defect-density fluctuations that substantially cancel in the total density response. We show that the long-wavelength fluctuation amplitudes agree with hydrodynamic theory, which is specified entirely by the elastic coefficients of the supersolid. Our analysis provides a microscopic connection between Bogoliubov excitations and the crystalline  and superfluid degrees of freedom of two-dimensional supersolids, and identifies their signatures in density- and current-sensitive probes.
\end{abstract}
\maketitle
%\tableofcontents
\section{Introduction}

 Two-dimensional (2D) supersolids have been realized in dipolar Bose--Einstein condensates (BECs) confined in oblate trapping geometries \cite{Norcia2021a,Bland2022a}, following the earlier realization of one-dimensional (1D) dipolar supersolids \cite{Tanzi2019a,Bottcher2019a,Chomaz2019a,Natale2019a,Guo2019a,Tanzi2019b}.
These systems exhibit a rich phase diagram, with several distinct crystalline
ground states separated by first-order phase transitions
\cite{Zhang2019a,Poli2021a,Hertkorn2021b,Zhang2021a,Zhang2023a,Ripley2023a,Lima2025a,Cook2026a}.
A supersolid combines spontaneous crystalline order with superfluidity
\cite{Andreev1969a,Leggett1970a,Chester1970a,Leggett1998a}.
The associated breaking of continuous translational and global \(U(1)\)
symmetries gives rise to three gapless Nambu--Goldstone excitation bands in
two dimensions \cite{Watanabe2012a}. Hydrodynamic theories accurately predict
the three corresponding speeds of sound
\cite{Rakic2024a,Poli2024b,Blakie2025a,Cook2026a}.

One branch is a transverse acoustic mode, whose existence demonstrates the
shear rigidity of the supersolid crystal. This mode has motivated proposals
for its excitation and detection
\cite{Yapa2025a}.
The remaining two acoustic branches are longitudinal modes that describe
coupled superfluid and crystalline motion. Methods for characterizing their
density fluctuations have been developed for 1D supersolids
\cite{Sindik2024a,Platt2024a,Mukherjee2025a,Bougas2026a}. Recent experiments on a driven BEC have measured
longitudinal sound propagation and crystal dynamics in a supersolid-like
system \cite{Liebster2025a}. However, the response to a transverse perturbation
was diffusive rather than a propagating transverse sound wave.

\begin{figure}[htbp!]
 \includegraphics[width=3.0in]{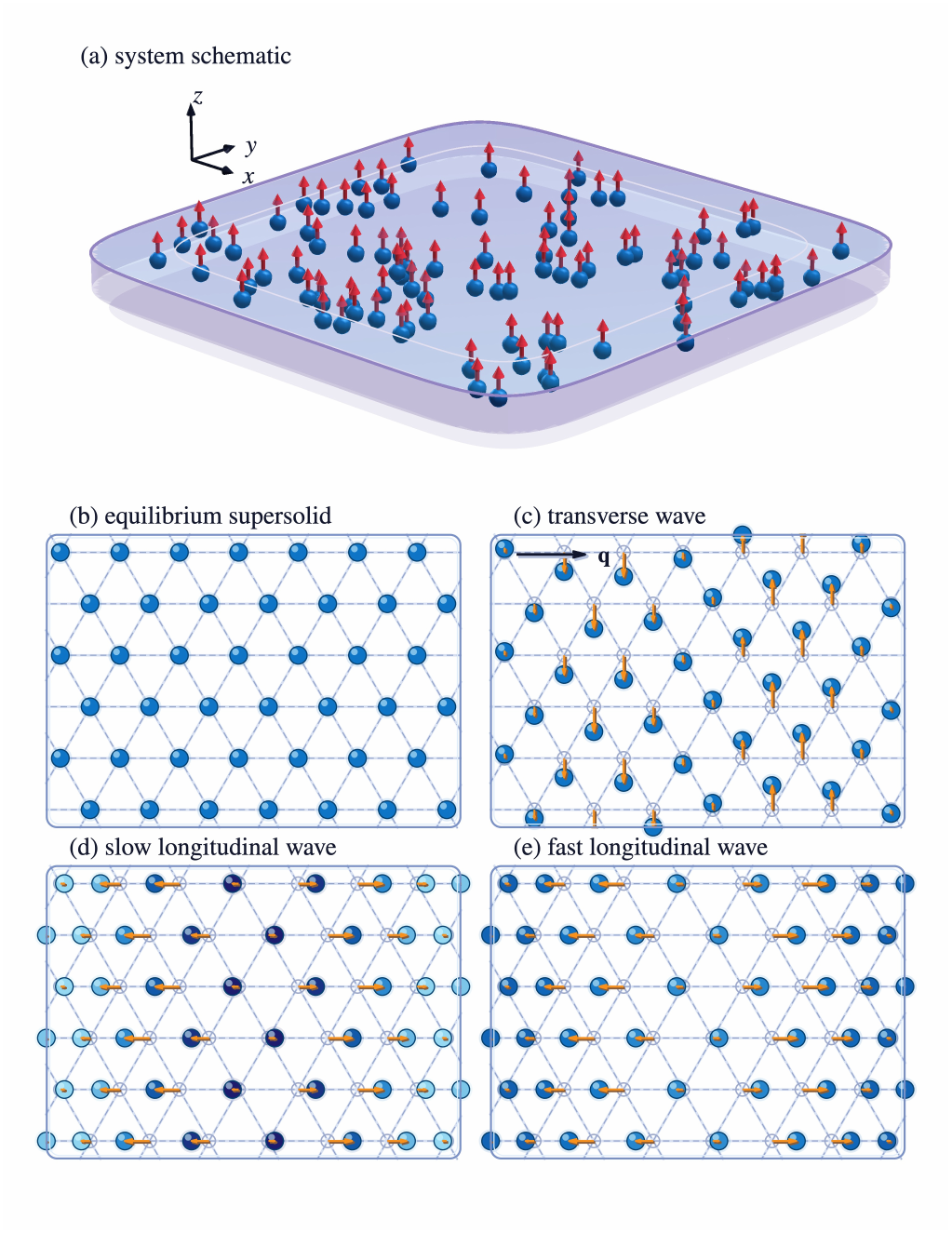} 
 \caption{Schematic of the system and its acoustic modes. (a) A dipolar Bose gas confined to a planar geometry, with the dipoles polarized along the harmonically confined $z$ direction. (b) An unperturbed triangular supersolid. (c)–(e) Schematics of the transverse, and two longitudinal acoustic modes. The arrows show the lattice displacement, while the site colour indicates the relative atom number at each site.}\label{figschematic}
\end{figure}

A detailed microscopic description of fluctuations in 2D supersolids remains
largely unexplored. In this work, we study a zero-temperature dipolar BEC
confined along \(z\), with dipoles polarized perpendicular to the
\(xy\) plane [see Fig.~\ref{figschematic}(a)] \cite{Baillie2015a}. We focus on
triangular and honeycomb supersolids, which are isotropic in their
long-wavelength sound speeds, superfluid response, and elastic properties
\cite{Zhang2019a,Ripley2023a}. Although we consider a dipolar system, the
microscopic analysis developed here should also apply to other 2D supersolids
for which a periodic ground state and its collective excitations can be
calculated, including soft-core supersolids
\cite{Rakic2024a,Poli2024b}.

We calculate the density, phase, and current fluctuations associated with the
acoustic Bogoliubov excitations. We further develop a procedure for extracting
the lattice displacement field directly from the microscopic density
perturbation. This allows the modes to be identified as one transverse and two
longitudinal branches, and enables the density fluctuations to be decomposed
into contributions from crystal strain and particle transport relative to the
lattice [see Figs.~\ref{figschematic}(b)-(e)]. These mode-resolved fluctuations characterize the coupled elastic and
superfluid dynamics of the supersolid and determine its response to weak
density-, phase-, and current-sensitive probes. We do not consider the 1D
stripe phase, whose collective dynamics are described by existing 1D
supersolid and generalized-smectic theories
\cite{Hofmann2021a,Sindik2024a,Platt2024a,Cook2026a,Poli2026a}.

The remainder of the paper is organized as follows. In
Sec.~\ref{Sec:Formalism}, we introduce the microscopic theory based on the
extended Gross--Pitaevskii equation (eGPE) and the associated
Bogoliubov--de Gennes (BdG) formalism. We also define the density, phase,
current, and displacement fluctuations. In Sec.~\ref{Sec:Results}, we present
the microscopic fluctuation results for triangular and honeycomb supersolids.
In Sec.~\ref{Sec:hydrosumruleresponse}, we connect their long-wavelength
limits to hydrodynamic theory, the elastic coefficients, and density and
current sum rules. Section~\ref{Sec:concl} presents our conclusions.

\section{Microscopic framework}  \label{Sec:Formalism}
\subsection{Ground states and phase diagram}
Here we discuss the details of our model of a planar dipolar BEC of magnetic atoms schematically shown in Fig.~\ref{figschematic}(a).
The magnetic dipole moments of the atoms are polarized along $z$ by a bias field and have the interaction  potential 
\begin{equation}
	U(\mathbf{r}) = \frac{4\pi a_s\hbar^2}{m}\delta(\br) + \frac{3\add\hbar^2}{m r^3}\left(1-3\frac{z^2}{r^2}\right),
\end{equation}
where $ \mathbf{r}= \mathbf{x}- \mathbf{x}'$ is the relative separation between the particles. Here $a_s$ is the $s$-wave scattering length,  $\add = m\mu_0\mu_m^2/12\pi\hbar^2$ is the dipole length, and $\mu_m$  is the atomic magnetic moment.
Stationary state solutions  $\Psi_0(\mathbf{x})$ for the condensate wavefunction can be taken as real and non-negative, and  satisfy the time-independent eGPE $\mathcal{L}\Psi_0=\mu\Psi_0$, where
\begin{align} 
\mathcal{L}&\equiv  -\frac{\hbar^2}{2m}\nabla^2 + \frac12 m\omega_z^2z^2 + \Phi   +\gammaQF\Psi_0^3. 
\end{align}
Here  $\Phi(\bx)=\int d\bx'\,U(\bx-\bx')|\Psi_0(\bx')|^2$,  with
 $\mu$ being the chemical potential and $\omega_z$ being the angular frequency of axial confinement.  The effects of quantum fluctuations are described by the term with coefficient $\gammaQF = \frac{128\pi\hbar^2}{3m}a_s\sqrt{\frac{a_s^3}{\pi}}\mathcal{Q}_5(\add/a_s)$, where $\mathcal{Q}_5(x)=\Re\{\int_0^1 du[1+x(3u^2 - 1)]^{5/2}\}$ \cite{Lima2011a,Ferrier-Barbut2016a,Wachtler2016a,Bisset2016a}. 
 Experimentally such a system can be prepared in a pancake-shaped harmonic trap   \cite{Norcia2021a,Bland2022a} or a box-shaped trap \cite{Navon2021a,Juhasz2022a}. Here we consider a uniform system in the $xy$-plane, but constrained to have an average areal density $\rho $.  This has the feature that excitations are translationally invariant and can be identified with well-defined quasimomentum.
  
\begin{figure}[htbp]
 \includegraphics[width=3.3in]{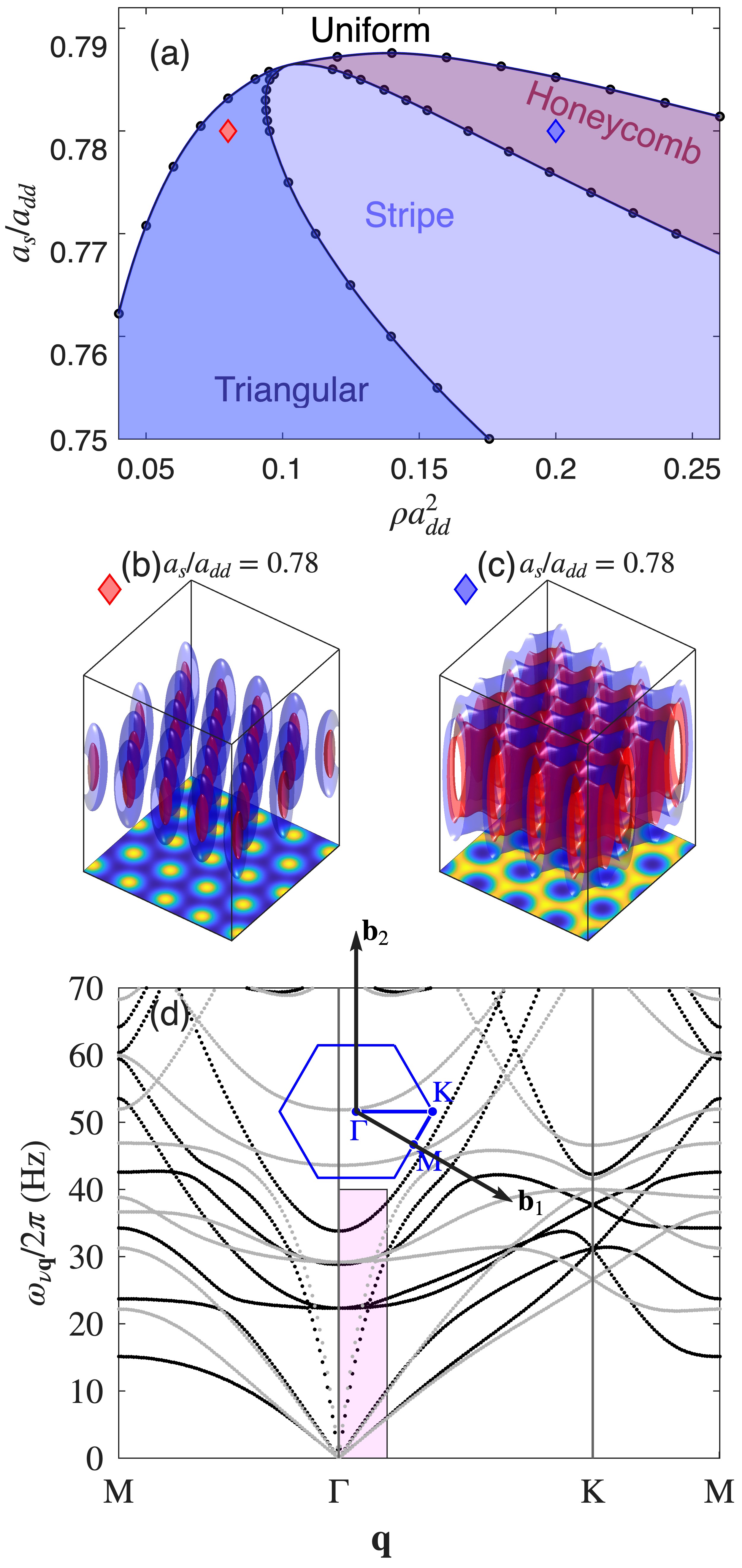}
\caption{Phase diagram, ground states and band structure for a planar dipolar BEC. (a) Phase diagram showing the regions where the uniform superfluid,  stripe crystal, triangular crystal and honeycomb crystal are the ground states.   Examples of (b) triangular [$\rho a_{dd}^2=0.08$] and (c) honeycomb [$\rho a_{dd}^2=0.2$]  ground states. The plot box is of size $20\!\times\!20 \times\!28\,\mu$m and the red (blue) isosurfaces are at a density of $3\times10^{20}\,$m$^{-3}$ ($1.5\times10^{20}\,$m$^{-3}$). A column density is given on the base plane. (d) Band structure of the triangular (black dots) and honeycomb (grey dots) supersolids corresponding to (b) and (c), respectively. Band structure shown along special directions of the first Brillouin zone, as shown in the inset. The shaded box indicates the low-$q$ region where we examine the properties of the long-wavelength excitations in this paper. Results for $^{164}$Dy with $a_{dd}=130.8\,a_0$ and $\omega_z/2\pi=72.4\,$Hz.}\label{figPD}
\end{figure}

 When $\add/a_s$ is sufficiently high, the ground state solution can develop 2D crystalline structure in the $xy$-plane \cite{Kadau2016a,Baillie2018a}, with direct lattice vectors of $\mathbf{a}_1$ and $\mathbf{a}_2$ defining the 2D unit cell. 
  In the results we present here we consider a system of $^{164}$Dy atoms in a harmonic trap of $\omega_z/2\pi=72.4\,$Hz,   similar to the parameters used for the phase diagrams presented in Refs.~\cite{Zhang2019a,Ripley2023a,Zhang2023a} and states explored in Refs.~\cite{Zhang2024a,Poli2024b,Blakie2024a}. 
The phase diagram is shown in Fig.~\ref{figPD}(a). At low densities the system has a triangular ground state, and at high densities it instead has a honeycomb ground state.  For both cases the direct lattice is hexagonal with the primitive vectors
\begin{align}
\mathbf{a}_1=a\hat{\mathbf{x}},\quad
\mathbf{a}_2=\frac{a}{2}\hat{\mathbf{x}}+\frac{\sqrt{3}a}{2}\hat{\mathbf{y}},\label{a12}
\end{align}
with $a$ being the lattice constant. In determining the ground state solution we also optimise the value of $a$ to minimise the energy per particle.
Here we illustrate typical triangular and honeycomb ground states in Figs.~\ref{figPD}(b) and (c).

\subsection{Excitations}
 The quasiparticle excitation modes $\{u_{\nu \mathbf{q}}(\mathbf{x}),v_{\nu \mathbf{q}}(\mathbf{x})\}$ and energies $\{\hbar\omega_{\nu \mathbf{q}}\}$ of the supersolid are determined by solving the BdG equations
\begin{align}
\begin{pmatrix} 
      \mathcal{L}+X-\mu & -X \\
      X & -(\mathcal{L}+X-\mu) \\
   \end{pmatrix} \begin{pmatrix} u_{\nu\mathbf{q}} \\ v_{\nu\mathbf{q}}  \end{pmatrix} =\hbar\omega_{\nu\mathbf{q}}\begin{pmatrix} u_{\nu\mathbf{q}} \\ v_{\nu\mathbf{q}}  \end{pmatrix},
\end{align}
where $\nu$ is the band index and $\mathbf{q}=(q_x,q_y)$ is a planar quasimomentum, which we restrict to lie in the first Brillouin zone.
Here $u_{\nu\bq}$ and $v_{\nu\bq}$ have the Bloch form, i.e.,~$u_{\nu\bq}(\bx)=\bar{u}_{\nu\bq}(\bx)e^{i\bq\cdot\bm{\rho}}$ with $\bar{u}_{\nu\bq}(\bx)$ periodic in the unit cell, and the exchange operator $X$ is defined so that 
\begin{align} 
\!Xf&=\!\Psi_0 \!\int\!d\mathbf{x}^\prime U(\mathbf{x}\!-\!\mathbf{x}^\prime)f(\mathbf{x}^\prime)\Psi_0(\mathbf{x}^\prime)  +\frac32\gamma_{\mathrm{QF}}\Psi_0^3f. 
\end{align}
The quasiparticles are normalized as $\int d\mathbf{x}\,[|u_{\nu\bq} |^2-|v_{\nu\bq} |^2]=1$ with the integration over all $z$ and the quantisation area of $A$. For the results in this paper we consider a system of 100 cells in a right-rhombic-prism normalization region with periodic boundary conditions on the $xy$-plane boundaries. This defines the condensate atom number as $N=\rho A$ removing ambiguity in the definition of the quasiparticle amplitude. We use a unit cell to solve for the ground state and excitations, using a continuous set of $\mathbf{q}$ values to map out the bands. General details about the numerical solution of these equations are discussed in Refs.~\cite{Poli2024b,Cook2026a}.

In Fig.~\ref{figPD}(d) we show examples of the band structure for triangular and honeycomb states, corresponding to ground states  in Figs.~\ref{figPD}(b) and (c). The band structure is shown along symmetry lines of the first Brillouin zone. This is defined by the reciprocal lattice vectors  
\begin{align}
\mathbf{b}_1=\frac{2\pi}{a}\hat{\mathbf{x}}-\frac{2\pi}{\sqrt{3}a}\hat{\mathbf{y}},\quad
\mathbf{b}_2= \frac{4\pi}{\sqrt{3}a}\hat{\mathbf{y}},\quad
\end{align}
with symmetry points  $\Gamma=\mathbf{0}$, $\mathrm{K}=\frac{2}{3}\mathbf{b}_1+\frac{1}{3}\mathbf{b}_2$, and $\mathrm{M}=\frac{1}{2}\mathbf{b}_1$ [see inset to Fig.~\ref{figPD}(d)]. 
In these results we see that there are three gapless excitation bands emerging at from the $\Gamma$ point [see shaded box in Fig.~\ref{figPD}(d)],  revealing the broken continuous symmetries \cite{Watanabe2012a}.

\subsection{Overview of excitation analysis}

 We will be interested in the evolution of various continuous fields $Q$ obtained from the system wavefunction as
 \begin{align}
 Q(\bm{\rho},t)=\int dz\,\mathcal{Q}(\Psi), \label{Qden}
 \end{align}
 where  $\bm{\rho}=(x,y)$ denotes the planar coordinate.   
 In this work we consider the following fields $Q\to\{\varrho,\vartheta,\mathbf{J}\}$, where
\begin{align}
\varrho (\bm{\rho},t)&=\int dz\,|\Psi (\mathbf{x},t)|^2,\label{dvarrho}\\
\vartheta (\bm{\rho},t)&=\int dz\,\mathrm{Arg}\{\Psi (\mathbf{x},t)\}\delta(z),\label{vartheta}\\
\mathbf{J} (\bm{\rho},t)&=  \int dz\, \frac{\hbar}{m}\, \mathrm{Im} \left[ \Psi ^*(\mathbf{x}, t)\, \bm{\nabla}_{\!\bm{\rho}} \Psi(\mathbf{x}, t) \right],\label{Jcur}
\end{align}
are the planar density, the in-plane phase, and planar current, respectively. The phase is undefined where the density vanishes, so we have defined $\vartheta$ on the $z=0$ plane. From these quantities we see that the  functional densities $\mathcal{Q}$  will in general depend on both the field and its gradients. 

 Our interest is on the effect of long-wavelength sound wave excitations on these fields. To do this we add a quasiparticle, with quantum numbers  $(\nu\,\mathbf{q})$, and an  amplitude of $c_{\nu\mathbf{q}}$ to the ground state, where $\nu=0,1,2$ and $|\mathbf{q}|a\ll1$ (i.e.,~long wavelength sound modes). For $|c_{\nu\mathbf{q}}|^2\ll N$  the evolution of the condensate field is well-described by
 \begin{align}
{\Psi}_{\nu\mathbf{q}}(\mathbf{x},t)=&e^{-i\mu t/\hbar}\left[\Psi_0+c_{\nu\mathbf{q}}u_{\nu\mathbf{q}} e^{-i\omega_{\nu\mathbf{q}}t} - c_{\nu\mathbf{q}}^* v^*_{\nu\mathbf{q}} e^{i\omega^*_{\nu\mathbf{q}}t}
\right].\label{psipert}
 \end{align}
 The effect of adding a quasiparticle is to create a wave in the $Q$-field of the form\footnote{Note this describes the coarse-grained  behavior of the field  by vitue of the projection onto the quasimomentum $\mathbf{q}$ (see discussion in Ref.~\cite{Platt2024a}).}
   \begin{align}
 \delta Q(\bm{\rho},t)=c_{\nu \mathbf{q}}A^{-1}\delta\tilde{Q}_{\nu \mathbf{q}}e^{i(\mathbf{q}\cdot\bm{\rho}-\omega_{\nu\mathbf{q}}t)}+\mathrm{c.c}. \label{Qres}
 \end{align}
We can obtain this result by linearizing Eq.~(\ref{Qden}) to first order in $c_{\nu\mathbf{q}}$ using  the expansion (\ref{psipert}), and then finding the amplitude of the Fourier component at wavevector $\mathbf{q}$.   
For the  fields of interest,   $\delta\tilde{Q}_{\nu\mathbf{q}} $  is given by the matrix elements
\begin{align}
\delta\tilde{\varrho}_{\nu\mathbf q} &= \int d\mathbf{x}\,\Psi_0(\mathbf{x})\bigl[u_{\nu\mathbf q}(\mathbf{x}) - v_{\nu\mathbf q}(\mathbf{x})\bigr]e^{-i\mathbf{q}\cdot \bm{\rho}},\label{drhot}\\
\delta\tilde{\vartheta}_{\nu\mathbf q}  &=  \int d\bm{\rho}\frac{u_{\nu\mathbf q}(\bm\rho,0) +v_{\nu\mathbf q}(\bm\rho,0)}{2i\Psi_0(\bm\rho,0)}e^{-i\mathbf{q}\cdot \bm{\rho}},\label{dthetat}\\
\delta\tilde{\mathbf{J}}_{\nu\mathbf q}  &= \frac{\hbar}{2mi}\int d\mathbf{x}\,\Bigl\{\Psi_0(\mathbf{x})\bm\nabla_{\bm\rho}\left[u_{\nu\mathbf q}(\mathbf{x}) + v_{\nu\mathbf q}(\mathbf{x})\right] \nonumber\\ 
&-\left[u_{\nu\mathbf q}(\mathbf{x}) + v_{\nu\mathbf q}(\mathbf{x})\right]\bm\nabla_{\bm\rho}\Psi_0(\mathbf{x})\Bigr\}e^{-i\mathbf{q}\cdot \bm{\rho}}.\label{dJt}
\end{align}
We will present results for these quantities in Sec.~\ref{Sec:Results}.  

\subsection{Crystal dynamics}\label{Sec:Xtalmotion} 
In general the low-energy excitations cause the crystalline structure to be perturbed, such that the  lattice sites are displaced in a time-varying manner. This displacement is naturally a discrete quantity, i.e.,~specific to each lattice site, but it can be described as a continuous field that we introduce below. 

 \begin{figure*}[htbp]
 \includegraphics[width=7in]{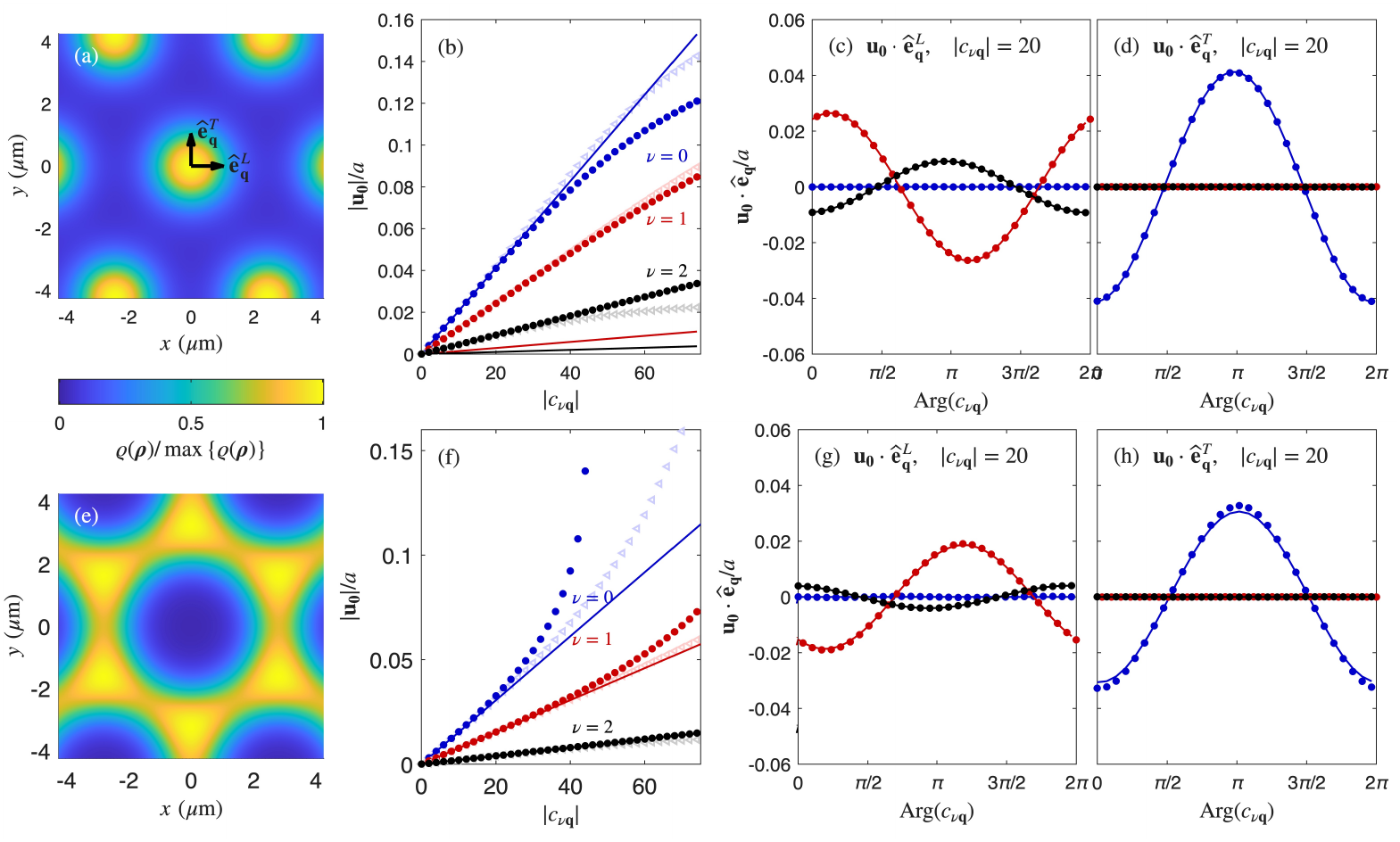}
\caption{Lattice site displacement for triangular (a)-(d) and honeycomb (e)-(h) supersolids caused by adding an excitation from the $\nu=0,1,2$  bands with a quasimomentun along $x$. 
(a), (e) show the unperturbed triangular and honeycomb areal densities $\varrho_0(\bm{\rho})$ and in (a) the longitudinal and transverse unit vectors are shown for reference. (b), (f) show the magnitude of the displacement of the site as a function of the quasiparticle amplitude. (c), (g) longitudinal and (d), (h) transverse displacement as a function of the phase of the quasiparticle.  Results in (b)-(d), (f)-(h) are coloured according to the band as labelled in (b) and (f).  Displacements computed from the first order expression for the density (\ref{perturbedden}) (dark circle markers) or from the full density  [i.e.,~including the terms second order in $c_{\nu\mathbf{q}}$ neglected in Eq.~(\ref{perturbedden})] (light triangle markers). The  linearized expression (\ref{uj})  is shown as a line.   Results in (a)-(d) are for $a_s/a_{dd}=0.77$, $\rho a_{dd}^2=0.08$, $\mathbf{q}=(0.026,0)\,\mu$m$^{-1}$ and $N=3.5\times10^6$. Results in (e)-(h) are for $a_s/a_{dd}=0.77$,  $\rho a_{dd}^2=0.20$, $\mathbf{q}=(0.022,0)\,\mu$m$^{-1}$ and $N=1.1\times10^7$. Other parameters as in Fig.~\ref{figPD}.
}\label{figdisplacement}
\end{figure*}

\subsubsection{Lattice site identification} 
 
We begin by identifying the equilibrium lattice-site locations. For the triangular supersolid, the lattice sites are naturally defined as the planar positions where the areal density attains its maxima within each unit cell [e.g., this occurs at the origin for the system shown in Fig.~\ref{figdisplacement}(a)]. In the regime considered here, lattice distortions remain small compared with the lattice constant, so that the density maximum associated with each site remains unique. In contrast, the honeycomb supersolid exhibits multiple density maxima within each unit cell, and it is therefore convenient to associate the lattice sites with the unique density minima [occurring at the origin for the example state shown in  Fig.~\ref{figdisplacement}(e)]. 

We denote the lattice position of the unperturbed ground state in the unit cell indexed by ${\mathbf{j}}$ as $\bm{\rho}_{\mathbf{j}}^0$, where  $\mathbf{j}=\{j_1,j_2\}$ is the Bravais-cell index, with  $j_1,j_2\in\mathbb{Z}$. For both the triangular and honeycomb states  $\{\bm{\rho}_{\mathbf{j}}^0\}$ forms a hexagonal Bravais lattice of points
\begin{align}
\bm{\rho}_{\mathbf{j}}^0=j_1\mathbf{a}_1+j_2\mathbf{a}_2,
\end{align}
where we have chosen to have the central site at the origin, i.e.,~$\bm{\rho}_\mathbf{0}^{0}=\mathbf{0}$.
 
\subsubsection{Displacement field}\label{Sec:dispfield}
The presence of a quasiparticle causes the lattice-site positions to shift in time to $\bm{\rho}_{\mathbf{j}}(t)$. The corresponding displacement vectors are defined as
\begin{align}
\mathbf{u}_{\mathbf{j}}(t)=\bm{\rho}_{\mathbf{j}}(t)-\bm{\rho}_{\mathbf{j}}^0 .
\end{align}
To determine these displacements we analyze the perturbed areal density
\begin{align}
\varrho(\bm{\rho},t)=\varrho_0(\bm{\rho})+\left[c_{\nu\mathbf{q}}\,\delta\varrho_{\nu\mathbf{q}}(\bm{\rho})e^{-i\omega_{\nu\mathbf{q}}t}+\mathrm{c.c.}\right],
\label{perturbedden}
\end{align}
where
\[
\varrho_0=\int dz\,|\Psi_0|^2,
\qquad
\delta\varrho_{\nu\mathbf q}=\int dz\,\Psi_0\bigl(u_{\nu\mathbf q}-v_{\nu\mathbf q}\bigr).
\]
Because the excitation carries quasimomentum $\mathbf q$, the density perturbation has Bloch form
\begin{align}
\delta\varrho_{\nu\mathbf q}(\bm{\rho})
= e^{i\mathbf q\cdot\bm{\rho}}\,
\delta\bar{\varrho}_{\nu\mathbf q}(\bm{\rho}),
\end{align}
where $\delta\bar{\varrho}_{\nu\mathbf q}$ is lattice periodic.

To determine the location of the density stationary point at time $t$, we consider a given unperturbed lattice site $\bm{\rho}_{\mathbf{j}}^0$ and assume that the displacement is small compared with the lattice spacing. The lattice site is defined as the stationary point of the density, so the displacement $\mathbf{u}_{\mathbf{j}}$ is determined by
\begin{align}
\bm{\nabla}_{\bm{\rho}}
\varrho(\bm{\rho}_{\mathbf{j}}^0+\mathbf{u}_{\mathbf{j}},t)
=
\mathbf{0}.
\end{align}

Expanding the gradient of Eq.~(\ref{perturbedden}) about $\bm{\rho}_{\mathbf{j}}^0$ yields
\begin{align}
\bm{\nabla}_{\bm{\rho}}
\varrho(\bm{\rho}_{\mathbf{j}}^0+\mathbf{u}_{\mathbf{j}},t)
\approx
H(\bm{\rho}_{\mathbf{j}}^0)\mathbf{u}_{\mathbf{j}}
+
\left[
c_{\nu\mathbf q}e^{-i\omega_{\nu\mathbf q}t}
\bm{\nabla}_{\bm{\rho}}\delta\varrho_{\nu\mathbf q}(\bm{\rho}_{\mathbf j}^0)
+\mathrm{c.c.}
\right],
\end{align}
where $H(\bm{\rho}_\mathbf{j}^0)$ is the Hessian matrix (i.e., the second derivatives with respect to the planar coordinates) of $\varrho_0$ evaluated at $\bm{\rho}_\mathbf{j}^0$. The linear term in $\varrho_0$ vanishes because $\bm{\rho}_\mathbf{j}^0$ is a stationary point.
Solving for the displacement gives
\begin{align}
\mathbf{u}_{\mathbf{j}}(t)= -H^{-1}(\bm{\rho}_\mathbf{j}^0)
\left[c_{\nu\mathbf q}e^{-i\omega_{\nu\mathbf q}t}
\bm{\nabla}_{\bm{\rho}}\delta\varrho_{\nu\mathbf q}|_{\bm{\rho}=\bm{\rho}_\mathbf{ j}^0}
+\mathrm{c.c.}
\right].
\end{align}
Using the periodicity of $H$ and $\delta\bar{\varrho}_{\nu\mathbf q}$, we obtain
\begin{align}
\mathbf{u}_{\mathbf{j}}(t)=c_{\nu\mathbf{q}}A^{-1}\tilde{\mathbf{u}}_{\nu \mathbf{q}}
e^{i(\mathbf{q}\cdot\bm{\rho}_\mathbf{j}^0-\omega_{\nu\mathbf{q}}t)}
+\mathrm{c.c.},
\label{uj}
\end{align}
where
\begin{align}
\tilde{\mathbf{u}}_{\nu \mathbf{q}}=-A\,H^{-1}(\mathbf{0})\,\bm{\nabla}_{\bm{\rho}}
\delta{\varrho}_{\nu\mathbf q}\big|_{\bm{\rho}=\mathbf{0}} .
\end{align}

In the continuum limit we introduce the displacement field $\mathbf{u}(\bm{\rho},t)$ such that $\mathbf{u}(\bm{\rho}_\mathbf{j}^0,t)=\mathbf{u}_{\mathbf{j}}(t)$. Therefore we have [setting $\bm{\rho}_\mathbf{j}^0\to\bm{\rho}$ in Eq.~(\ref{uj})]
\begin{align}
\mathbf{u}(\bm{\rho},t)=c_{\nu\mathbf{q}}A^{-1}\tilde{\mathbf{u}}_{\nu \mathbf{q}}
e^{i(\mathbf{q}\cdot\bm{\rho}-\omega_{\nu\mathbf{q}}t)}
+\mathrm{c.c.}
\label{ut}
\end{align}   

\subsubsection{Displacement results}\label{Sec:DisplacementResults} 

In Fig.~\ref{figdisplacement} we analyze the displacement amplitude $\mathbf{u}_{\mathbf{0}}$ of the central lattice site for triangular [subplots (a)–(d)] and honeycomb [subplots (e)–(h)] supersolids. The results correspond to adding an excitation from each of the $\nu=0,1,2$ bands.
Subplots (b) and (f) show the magnitude of the displacement as a function of the excitation amplitude $|c_{\nu \mathbf{q}}|$. The linear result of Eq.~(\ref{ut}) is seen to hold for $|c_{\nu \mathbf{q}}|\lesssim 50$, corresponding to displacements $\lesssim 5\%$ of the lattice constant. For larger amplitudes the behavior becomes band dependent (e.g., the displacement may grow more rapidly or saturate). In this regime second-order contributions in $c_{\nu \mathbf{q}}$ become relevant.

Since the displacement is a vector quantity, it is convenient to decompose it into longitudinal and transverse components by introducing the polarization vectors
\begin{align}
\hat{\mathbf e}^{\mathrm L}_{\mathbf q} \equiv \frac{\mathbf q}{q},
\qquad
\hat{\mathbf e}^{\mathrm T}_{\mathbf q} \equiv
\hat{\mathbf z}\times\frac{\mathbf q}{q},
\end{align}
where $q\equiv|\mathbf q|$. This decomposition allows us to distinguish compressional (longitudinal) and shear (transverse) lattice responses.

We show the projection of the displacement onto these polarization vectors for the triangular supersolid [subplots (c) and (d)] and the honeycomb supersolid [subplots (g) and (h)]. In these plots the phase of $c_{\nu\mathbf{q}}$ is varied over $2\pi$, corresponding to the dynamical motion of the lattice site over one period of the excitation.
These results show that for both supersolids the lowest band ($\nu=0$) corresponds to a transversely polarized mode (i.e., a shear wave), while the $\nu=1$ and $\nu=2$ bands correspond to longitudinal modes.

\begin{figure}[htbp]
 \includegraphics[width=3.4in]{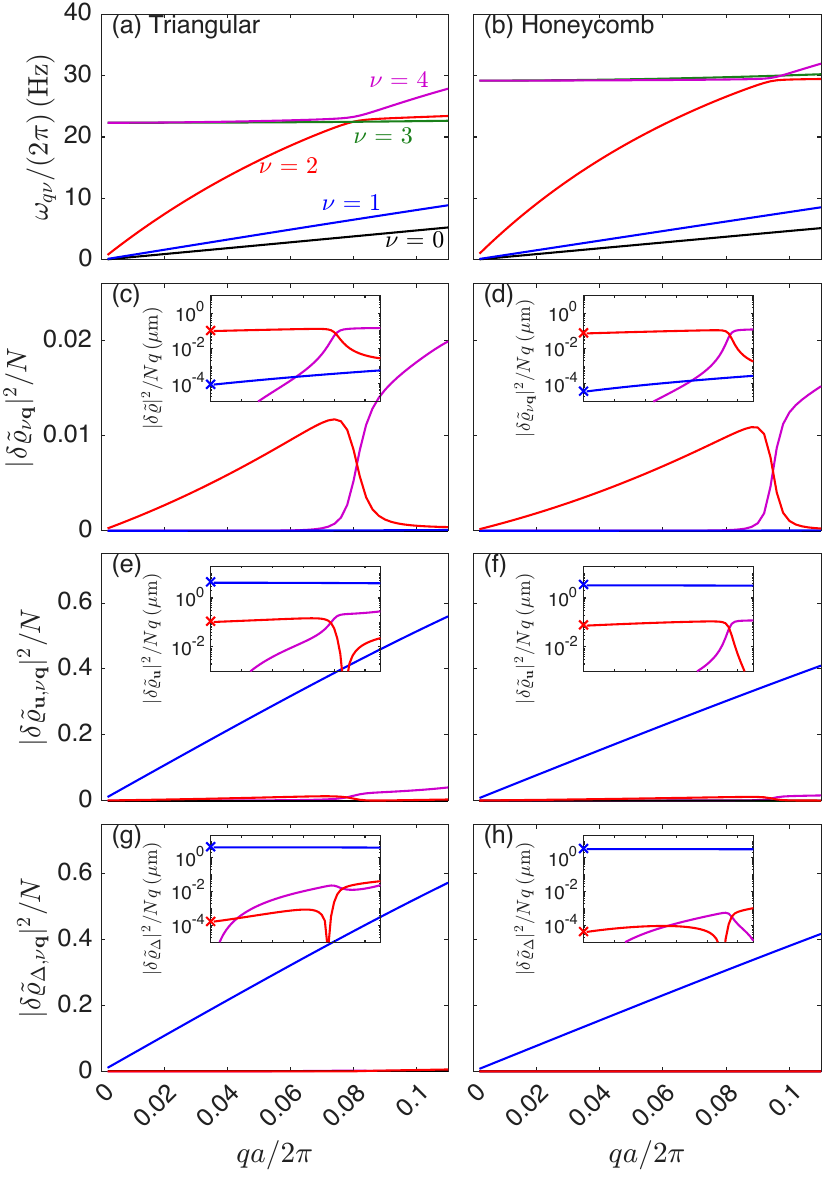}
\caption{Density fluctuations for triangular (left column) and honeycomb (right column) states. (a), (b) Excitation spectrum showing lowest 5 bands. (c), (d) Total density fluctuation $\delta\tilde{\varrho}_{\nu\mathbf{q}}$ [from Eq.~(\ref{drhot})]. (e), (f) strain density fluctuation $\delta\tilde{\varrho}_{\mathbf{u},\nu\mathbf{q}}$ [from Eq.~(\ref{deltau})] , and (g), (h) defect density fluctuation $\delta\tilde{\varrho}_{\Delta,\nu\mathbf{q}}$ [from Eq.~(\ref{deltaD})]. Insets to (c)-(h) show the fluctuation scaled by wave vector with the crosses indicating the hydrodynamic theory asymptotes. Both states are at $a_s/a_{dd}=0.780$ with a density of $\rho =0.08/a_{dd}^2$  ($\rho =0.20/a_{dd}^2$) for the triangular (honeycomb) case. The lattice constant is $a=4.93\,\mu$m ($a=5.38\,\mu$m)  for the triangular (honeycomb) case.  }\label{figdenflucts}
\end{figure}

\section{Fluctuation results}\label{Sec:Results}
In this section we examine the density fluctuations, first considering the total density fluctuations. Then, using our results for the displacement field, we quantify the component of the fluctuations arising from crystal strain. This allows us to infer the  defect-density fluctuations associated with particle transport relative to the lattice, i.e.,~arising from the tunnelling of atoms between sites in the supersolid.

\subsection{Total density fluctuations} 
The total density fluctuations $\delta\tilde{\varrho}_{\nu\mathbf{q}}$, shown in Figs.~\ref{figdenflucts}(c) and (d), are obtained from the numerical solution of the BdG equations using Eq.~(\ref{drhot}).  
At low $q$, the fluctuations are strongest for excitations in the $\nu=2$ band, with a much weaker contribution from the $\nu=1$ band. 
Subplots (a) and (b) show the related band structure, including an avoided crossing occurring as $q$ increases,  where density fluctuation weight transfers between the $\nu=2$ and $\nu=4$ bands.
The $\nu=0$ band does not contribute to the density fluctuations, consistent with the shear-wave character of this mode (see Sec.~\ref{Sec:DisplacementResults}). 

\subsection{Strain density fluctuations}
 The fractional change in area of a lattice site under strain is given by the divergence of the displacement field $\delta  A_{\mathrm{uc}}/A_{\mathrm{uc}}=\bm{\nabla}_{\bm{\rho}}\cdot\mathbf{u}(\bm{\rho},t)$. If the number of atoms at the lattice site is fixed,  then this change in area leads to a corresponding change in the density. We identify this with the strain-induced density fluctuation  
\begin{align}
\delta\varrho_{\mathbf{u}}(\bm{\rho},t)=-\rho \bm{\nabla}_{\bm{\rho}}\cdot\mathbf{u}(\bm{\rho},t).
\end{align}
Applying this expression to a $(\nu,\mathbf{q})$ quasiparticle using Eq.~(\ref{ut}), we obtain the strain density fluctuation of the form (\ref{Qres}), with amplitude
\begin{align}
\delta\tilde{\varrho}_{\mathbf{u},\nu\mathbf{q}}=-i\rho \mathbf{q}\cdot\tilde{\mathbf{u}}_{\nu\mathbf{q}}.\label{deltau}
\end{align}
This also shows that the density fluctuation only depends on the longitudinal component of the displacement field.

We present results for $\delta\tilde{\varrho}_{\mathbf{u},\nu\mathbf{q}}$ in Fig.~\ref{figdenflucts}(e) and (f). These results show a strong strain density fluctuation for the $\nu=1$ band and a weaker response from the $\nu=2$ band. The $\nu=0$ band does not contribute because the displacement field for this branch is transverse.

\subsection{Defect density fluctuations}
Defect density fluctuations are the difference between the total density fluctuations (\ref{drhot}) and the density fluctuations from crystal displacement (\ref{deltau})
\begin{align}
\delta\tilde{\varrho}_{\Delta,\nu \mathbf{q}}=\delta\tilde{\varrho}_{\nu \mathbf{q}}-\delta\tilde{\varrho}_{\mathbf{u},\nu \mathbf{q}}.\label{deltaD}
\end{align}
This arises because tunnelling between sites allows the number of atoms at each site of the supersolid to change.

Results for $\delta\tilde{\varrho}_{\Delta,\nu\mathbf{q}}$ are presented in Fig.~\ref{figdenflucts}(g) and (h). These results show a strong defect-density fluctuation for the $\nu=1$ band and a weaker response from the $\nu=2$ band. We note that while the amplitudes for the strain and defect densities are large for the $\nu=1$ band, they are out of phase and cancel out in the total density fluctuation.  Thus, when the lattice locally compresses, particles move out of the contracting cells, producing a defect-density fluctuation that largely compensates the strain-induced increase in density.

\begin{figure}[htbp]
 \includegraphics[width=3.4in]{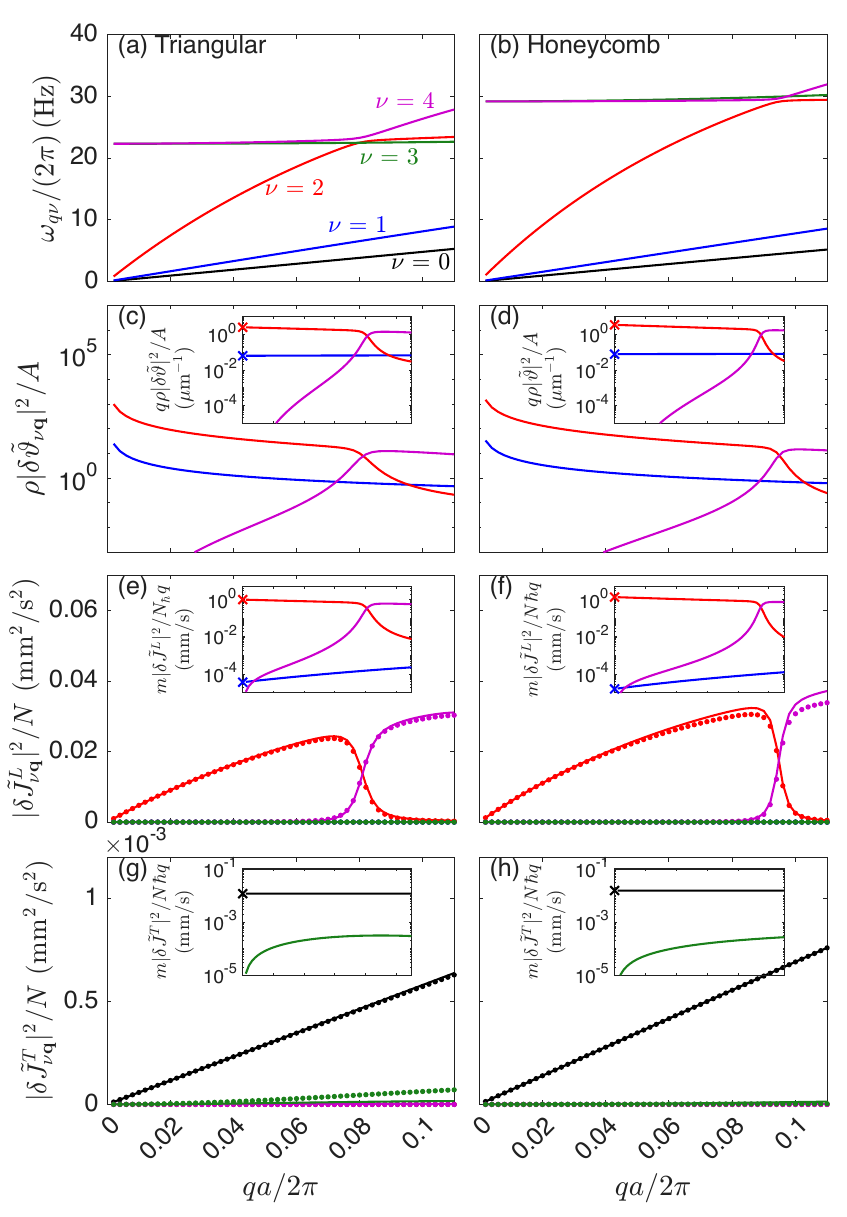}
\caption{Phase and current fluctuations for triangular (left column) and honeycomb (right column) states. (a), (b) Excitation spectrum showing lowest 5 bands. (c), (d) Phase fluctuations $\delta\tilde{\vartheta}_{\nu\mathbf{q}}$ [from Eq.~(\ref{dthetat})]. (e), (f) longitudinal current fluctuations $\delta\tilde{J}^L_{\nu\mathbf{q}}$  and (g), (h) transverse current fluctuations $\delta\tilde{J}^T_{\nu\mathbf{q}}$ [from Eq.~(\ref{dJt})]. Insets to (c)-(h) show the fluctuation scaled by wave vector with the crosses indicate the   hydrodynamic theory asymptotes. The dots in (e) - (h)   test the applicability of Eqs.~(\ref{Jfromphiu}) and (\ref{JfromuT}) [see text].  The states analyzed are the same as in Fig.~\ref{figdenflucts}.   }\label{figphasecurrentflucts}
\end{figure}

\subsection{Phase fluctuations}
 The phase fluctuations are generally closely related to the density fluctuations. However, while the density fluctuations tend to be suppressed as $q\to 0$, the phase fluctuations instead diverge [see Figs.~\ref{figphasecurrentflucts}(c) and (d)]. We see that the  $\nu=2$ band dominates the low $q$ fluctuations, with the $\nu=1$ band giving a small contribution. The transverse sound band $\nu=0$ has a vanishing contribution.
  
\subsection{Current fluctuations}
 The planar current fluctuation amplitude is a vector quantity and, as for the displacement field, we decompose it into longitudinal and  transverse components. The results in Figs.~\ref{figphasecurrentflucts}(e) - (h) show that the $\nu=1,2$ bands are longitudinal, and at low $q$ the $\nu=2$ band dominates. Similarly, the $\nu=0$ band is transverse and dominates in the $q\to0$ limit.

\section{Hydrodynamic limits and response}  \label{Sec:hydrosumruleresponse}
\subsection{Hydrodynamics fields}

In supersolid hydrodynamic theory \cite{Son2005a,Yoo2010a} (also see \cite{Andreev1969a,Saslow1977a,Liu1978a,Josserand2007a,Josserand2007b,Ye2008a,Hofmann2021a,Buhler2023a})
the long-wavelength dynamics of a  2D supersolid are governed by three coarse-grained fields: the areal density $\rho$, superfluid phase $\theta$, and lattice displacement $\mathbf{u}$. We note that: (i) we already defined $\rho=N/A$ as the average areal density, but in the hydrodynamic description this can vary slowly over space; (ii)  $\theta$ is the on-plane  phase obtained by coarse-graining $\vartheta$; (iii) the displacement field is already coarse-grained, since by its construction it contains no short wavelength structure (see Sec.~\ref{Sec:dispfield}).  

Global U(1) and translational invariance imply that the energy density $\mathcal{E}$ depends on the
phase and displacement only through the superfluid velocity $v_{s,i}=\frac{\hbar}{m}\partial_i\theta$, and the
strain tensor $u_{ij}=\frac{1}{2}(\partial_iu_j+\partial_ju_i)$, with $i,j=\{x,y\}$. 
Elastic parameters emerge by making a quadratic expansion of the energy density. 
In practice we can obtain the energy density on a single unit cell using the eGPE theory as $\mathcal{E}(\rho,\mathbf{v}_s,u_{ij})$, where we treat the coarse-grained fields as constants.  Finite differences in the values of these constants are then used to obtain the elastic parameters. The general procedure is discussed in detail in Ref.~\cite{Cook2026a}. In what follows we briefly introduce each of the elastic parameters and consider key predictions from hydrodynamic theory. 

\subsection{Superfluid density and current}
The superfluid density tensor is an elastic parameter which defines the response of the system to twists in phase, or equivalently an imposed superfluid velocity \cite{Leggett1970a,Saslow1976a,Sepulveda2010a,Biagioni2024a,Platt2025a,Perez-Cruz2025a}
\begin{align}
\rho_{s,ij}=\frac{1}{m}\frac{\partial^2\mathcal{E}}{\partial v_{s,i}\partial v_{s,j}}.
\end{align}  
This quantity is isotropic for the triangular and honeycomb states we consider \cite{Blakie2024a} with $\rho_{s,ij}=\rho_s\delta_{ij}$, with $\rho_s$ being the scalar superfluid density (also see \cite{Tao2023a,Chauveau2023a,Rabec2025a}). We can also introduce $\rho_n=\rho-\rho_s$ as the normal density\footnote{Here we follow the commonly used terminology in the literature, but note that this could also be referred to as the supersolid density $\rho_L$  \cite{Saslow2025a}.  Note that at $T=0$ this component is coherent and does not carry entropy \cite{Preti2026a}. }. An important prediction of supersolid hydrodynamics is that current in the system can be generally written in the Andreev-Liftshitz two-fluid form \cite{Andreev1969a}
 \begin{align}
 \mathbf J=\rho_s\mathbf v_s+\rho_n\mathbf v_n,\label{twofluid}
 \end{align}
 with the superfluid velocity related to the gradient of the coarse-grained phase field, and in the zero temperature regime we study here, the normal velocity can be related to the motion of the crystal as $\mathbf{v}_n\equiv\partial_t\mathbf{u}$.
 This imposes the following Fourier space relationships between the current, phase and displacement fluctuations
  \begin{align}
  \delta\tilde{J}_{\nu\mathbf{q}}^L&=iq\rho_s\frac{\hbar}{m}\delta\tilde{\vartheta}_{\nu\mathbf{q}}-i\omega_{\nu\mathbf{q}}\rho_n\tilde{u}_{\nu\mathbf{q}}^L,\label{Jfromphiu}\\
  \delta\tilde{J}_{\nu\mathbf{q}}^T&= -i\omega_{\nu\mathbf{q}}\rho_n\tilde{u}_{\nu\mathbf{q}}^T,\label{JfromuT}
  \end{align}
  where $\tilde{u}_{\nu\mathbf{q}}^L$ and $\tilde{u}_{\nu\mathbf{q}}^T$ are the longitudinal and transverse components of $\tilde{\mathbf{u}}_{\nu\mathbf{q}}$.
  We test these results in Figs.~\ref{figphasecurrentflucts}(e)-(h), where we evaluate the right hand side of Eqs.~(\ref{Jfromphiu}) and (\ref{JfromuT}) using our results for $\{\rho_n,\rho_s,\omega_{\nu\mathbf{q}},\tilde{\mathbf{u}}_{\nu\mathbf{q}}, \delta\tilde{\vartheta}_{\nu\mathbf{q}}\}$ and compare it to the current fluctuations calculated directly from the quasiparticles. While Eqs.~(\ref{Jfromphiu}) and (\ref{JfromuT}) are expected to hold in the $q\to0$ limit, our results show  that the relations work quite well over the entire $q$ range considered. For the $\nu=1$ band the superfluid and normal contributions to Eq.~(\ref{Jfromphiu}) are out of phase, leading to a reduction in the amplitude of the current fluctuations. For the $\nu=2$ band these two contributions are in phase.

\subsection{Asymptotic behavior of fluctuations}
To introduce the hydrodynamic results for the asymptotic behavior of the fluctuations we first introduce the remaining elastic parameters and the speeds of sound.
The lattice elastic tensor $C_{ijkl}$  is obtained as the second derivative with respect to the strain tensor $u_{ij}$  \cite{Bavaud1986a}.  For the triangular and honeycomb supersolids the elastic tensor is isotropic \cite{LandauElasticity} of the form
\begin{align}
C_{ijkl}=\tilde\lambda\delta_{ij}\delta_{kl}+\tilde\mu(\delta_{ik}\delta_{jl}+\delta_{il}\delta_{jk}),\label{Cijkl}
\end{align}
where $\{\tilde\lambda,\tilde\mu\}$ are the Lam\'e parameters, with $\tilde\mu$ being the shear modulus.  It is useful to introduce $\alpha_{uu}=C_{xxxx}=\tilde{\lambda}+2\tilde{\mu}$ as the longitudinal elastic modulus of the crystal.

Changes in density are described by the density stiffness
\begin{align}
\alpha_{\rho\rho} = \frac{\partial ^2\mathcal{E}}{\partial \rho^2},
\end{align}
and the coupling between density and diagonal-strain is characterized by the density-strain parameter 
\begin{align}
\alpha_{\rho u}= \frac{\partial ^2\mathcal{E}}{\partial \rho\partial u_{ii}},\qquad \mbox{(no summation on $i$)}.
\end{align}  
This term accounts for the coupling of changes in cell area (i.e.~normal strains $\{u_{xx},u_{yy}\}$)  to changes in density. Generally this elastic parameter is small, but is important for quantitative descriptions of the speeds of sound (see \cite{Platt2024a}).

The elastic parameters  determine the hydrodynamic expressions for the supersolid speeds of sound   \cite{Yoo2010a,Platt2024a,Poli2024b} as
\begin{align}
    mc_\pm^2&= \varepsilon  \pm \sqrt{ {\varepsilon^2} \!-\!  \frac{\rho_{s}}{\rho_{n}}\left(\alpha_{\rho\rho}\alpha_{uu}-\alpha_{\rho u}^2\right)},\\
  mc_T^2&={\frac{\tilde{\mu}}{\rho_{{n}}}}.\label{speedsofsound}
    \end{align} 
Here  $c_\pm$ and $c_T$ are the longitudinal and transverse sound speeds, respectively,
where $\varepsilon  =  (\rho\alpha_{\rho\rho}-2\alpha_{\rho u} +  {\alpha_{uu}}/{\rho_n})/2$.
 Comparisons between hydrodynamic and BdG calculations of the speed of sound have already been performed in Refs.~\cite{Blakie2023a,Platt2024a,Poli2024b,Rakic2024a,Blakie2025a,Cook2026a}  and show excellent agreement.
 In what follows for simplicity of notation we use $T$ to denote the transverse sound band ($\nu=0$), and $-$ for the lower ($\nu=1$) and $+$ for the upper ($\nu=2$) longitudinal sound bands.
 
The non-trivial asymptotic behavior of fluctuations for the longitudinal sound bands is
 \begin{align}
 {|\delta \tilde\varrho_{\pm q}|^2} &\underset{q\to0}{\sim}\frac{1}{2}Nq\xi_\rho^{\pm},\label{frho}\\
 {|\delta \tilde\varrho_{u,\pm q}|^2}&\underset{q\to0}{\sim}\frac{1}{2}Nq\xi_u^{\pm}, \label{fu}\\ 
 {|\delta \tilde\varrho_{\Delta,\pm q}|^2}&\underset{q\to0}{\sim}\frac{1}{2}Nq\xi_\Delta^{\pm},  \label{fDelta}\\ 
 |\delta\tilde\vartheta_{\pm q}|^2&\underset{q\to0}{\sim}\frac{A}{2\rho}\frac{1}{q\xi_\phi^{\pm}}, \label{ftheta}\\ 
 {|\delta\tilde{J}^L_{\pm q}|^2}&\underset{q\to0}{\sim}\frac{1}{2}N c_{\pm}^2{ q}\xi_\rho^{\pm}, \label{fJL} 
 \end{align}
 where we have introduced the correlation lengths\footnote{In the uniform superfluid $\rho_n\to0$ (i.e.~$\rho_s\to\rho$), and there is a single longitudinal supersolid sound band, with  speed $c =\sqrt{\rho\alpha_{\rho\rho}/m}$. Here the  non-zero correlation lengths are $\xi_\rho\to\xi $ and $\xi_\phi\to\xi $, with $\xi =\hbar/mc$.}
 \begin{align}
  \xi_\rho^{\pm} &=  \frac\hbar{m\rho}\frac{\rho m c_\pm^2-\frac{\rho_s}{\rho_n}\alpha_{uu}}{mc_\pm(c_\pm^2-c_\mp^2)},\label{lrho}\\
    \xi_{u}^{\pm} &=  \frac{\hbar\rho}{m\rho_n}\frac{ m c_\pm^2-\rho_s\alpha_{\rho\rho}}{mc_\pm(c_\pm^2-c_\mp^2)},\\
  \xi_{\Delta}^{\pm} &=  \frac{\hbar\rho_s}{m\rho\rho_n}\frac{\rho m c_\pm^2-(\alpha_{uu} - 2\rho\alpha_{\rho u} + \rho^2\alpha_{\rho\rho})}{mc_\pm(c_\pm^2-c_\mp^2)}, \\
 \xi_\phi^{\pm}  &=   \frac{\hbar}{\rho}
\frac{mc_\pm(c_\pm^2-c_\mp^2)}{\alpha_{\rho\rho} m c_\pm^2-\frac{1}{\rho_n}(\alpha_{\rho\rho}\alpha_{uu}-\alpha_{\rho u}^2)}.
  \end{align}
Results (\ref{frho})-(\ref{ftheta}) were previously derived and used in application to 1D supersolids in Ref.~\cite{Platt2024a}, which also has two longitudinal sound modes.  Result (\ref{fJL})  follows immediately from the density fluctuation result [Eqs.~(\ref{frho}) and (\ref{lrho})] using the continuity equation in Fourier space  $\omega\,\delta  \varrho =\mathbf{q}\cdot\delta\mathbf{J}$.

The transverse sound band only contributes to the transverse current fluctuations
\begin{align}
 {|\delta\tilde{J}^T_{0q}|^2}\underset{q\to0}{\sim}\frac{1}{2}Nqc_T^2\xi_T , \label{JTphi}
\end{align}
where 
\begin{align}
\xi_T =\frac{\rho_n}{\rho}\frac{\hbar}{mc_T}.\label{xiT}
\end{align}
%Equation~(\ref{deltau}) maps longitudinal displacement to strain density, whereas Eq.~(\ref{JfromuT}) maps transverse displacement to tranverse current.
This result can be obtained from the sum rule (\ref{transcurrule}) we discuss in the next subsection, using that the transverse mode dominates the contribution to the transverse current fluctuations in the small $q$ limit. 
%We previously mentioned the connection between the strain density fluctuations and the longitudinal displacement field [Eq.~(\ref{deltau}); also see Eq.~(\ref{JfromuT})].
 We confirm results (\ref{frho})-(\ref{fJL}) and (\ref{JTphi}) in the insets of Figs.~\ref{figdenflucts} and \ref{figphasecurrentflucts}, where the cross markers are used to indicate the $q\to0$ behavior. The values of the elastic parameters and speeds of sound for the states analyzed here are given in Table \ref{tab:hydrodynamic_parameters}.
 
 \begin{table}[t]
\caption{
Hydrodynamic parameters for the two supersolid
states considered in Figs.~\ref{figdenflucts}-\ref{figsumrules}.
}
\label{tab:hydrodynamic_parameters}
\begin{ruledtabular}
\begin{tabular}{lccc}
Quantity & Units & Triangular & Honeycomb \\[1pt]
\hline 
\rule{0pt}{2.5ex}
  $\rho$             
    & $\mathrm{m}^{-2}$ & $1.67\times10^{15}$ & $4.17\times10^{15}$  \\[2pt] 
%\multicolumn{4}{c}{Elastic coeffs.} \\[1pt]
  $\rho_s/\rho$
    & $1$ & $0.900$ & $0.879$ \\
$\rho\alpha_{\rho\rho}/h$
    & $\mathrm{Hz}$ & $1.68\times10^3$ &  $3.13\times10^3$ \\
$\alpha_{\rho u}/h$
    & $\mathrm{Hz}$ & $0.336$ & $0.0904$ \\
$\alpha_{uu}/\rho h$
    & $\mathrm{Hz}$ & $7.67$ & $10.2$ \\
$\tilde{\mu}/\rho h$
    & $\mathrm{Hz}$ & $2.27$ & $3.23$ \\[2pt]
%\multicolumn{4}{c}{Sound speeds} \\[1pt]
$c_T$ (or $c_0$)
    & $\mathrm{mm\,s^{-1}}$ & $0.235$ & $0.254$ \\
$c_-$ (or $c_1$)
    & $\mathrm{mm\,s^{-1}}$ & $0.409$ & $0.422$ \\
$c_+ $ (or $c_2$)
    & $\mathrm{mm\,s^{-1}}$ & $2.03$ & $2.76$ \\
\end{tabular}
\end{ruledtabular}
\end{table}

\begin{figure}[htbp]
 \includegraphics[width=3.4in]{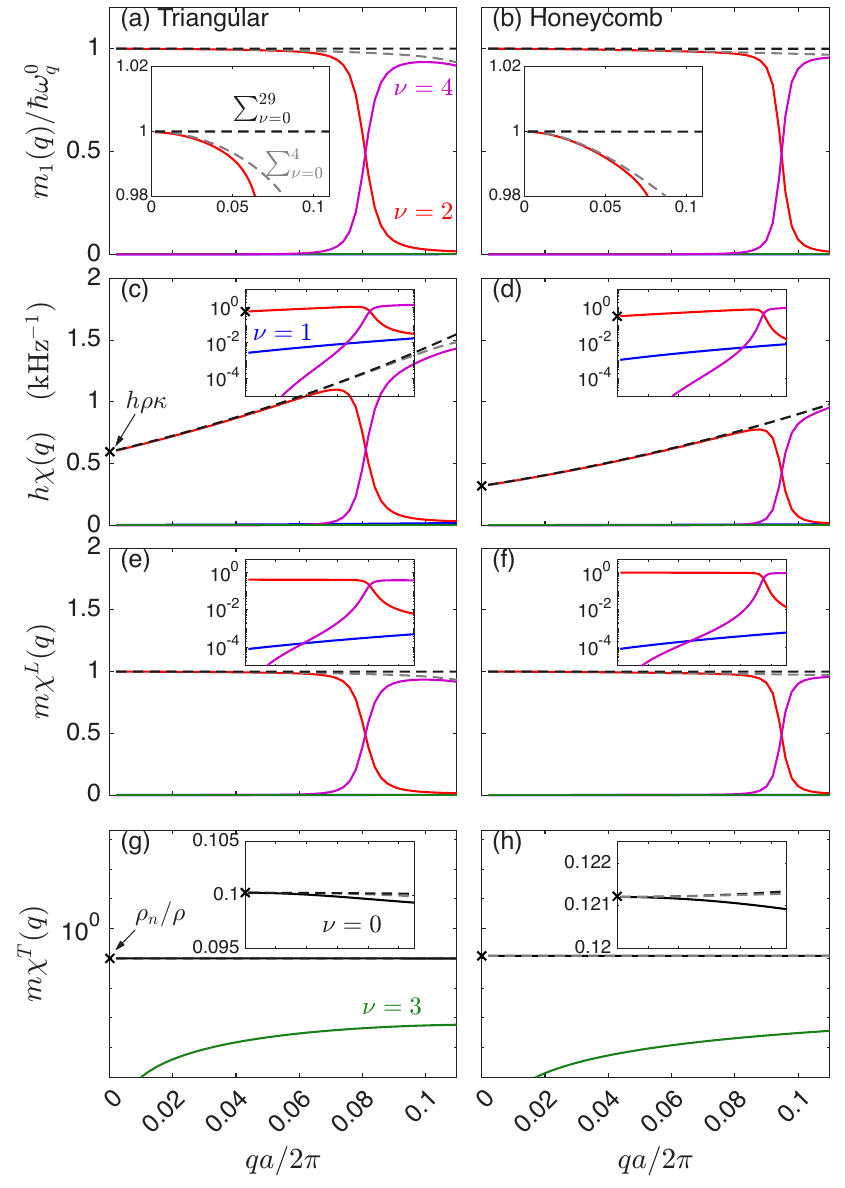}
\caption{Comparison of sum rules and response functions for triangular (left column) and honeycomb (right column) states. (a), (b) The first moment of the dynamical structure factor divided by the free particle energy.  (c), (d) Static density response function.   (e), (f) Longitudinal  and (g), (h) transverse current static response functions. Insets zoom in to regions of interest or reveal the behavior using logarithmic axes. Individual contributions of bands to the results are shown as colored lines as labelled. Results obtained by summing over the first 5 bands (grey dashed lines) or the first 30 bands (black dashed lines). The  crosses at $q=0$ in (c)-(d) and (g)-(h) indicate the macroscopic limits discussed in Sec.~\ref{Sec:sumrule}.  For (a)-(b) and (e)-(f) the sum rule is satisfied when the result is unity independent of $q$. The states analyzed are the same as in Fig.~\ref{figdenflucts}.   
  }\label{figsumrules}
\end{figure}

\subsection{Sum rules and response functions} \label{Sec:sumrule}
The dynamical structure factor determines the response of the system to a density-coupled probe (see \cite{Petter2021a,Bismut2012a,Petter2019a}), where the probe transfers momentum $\hbar\mathbf{q}$ and energy $\hbar\omega$. 
For a zero temperature BEC it is given by \cite{Zambelli2000,Blakie2002a} 
\begin{align}
S(\mathbf{q},\omega)=\sum_{\nu}|\delta\tilde{\varrho}_{\nu,\mathbf{q}}|^2\delta(\hbar\omega-\hbar\omega_{\nu\mathbf{q}}).
\end{align} 
We will be interested in the moments of this function defined as \cite{BECbook}
\begin{align}
m_p(\mathbf{q})=\hbar^{p+1}\int_{-\infty}^{+\infty}\omega^pS(\mathbf{q},\omega)d\omega.
\end{align}

The $p=1$ moment satisfies the $f$-sum rule \cite{PinesNozieres1966,BECbook}
\begin{align}
m_1(\mathbf{q})=N\hbar\omega^0_q,
\end{align}
where $\omega^0_q=\hbar q^2/2m$ is the free particle dispersion relation. We examine this sum rule in Figs.~\ref{figsumrules}(a) and (b), comparing the contributions from the lowest 5 branches, and the partial sum of the first 5 and first 30 branches. These reveal that at low $q$ the $\nu=2$ (upper longitudinal) sound band dominates. As $q$ increases we need to sum over more bands to ensure the sum rule is satisfied.

The $p=-1$ moment relates to the static density response function $N\chi(\mathbf{q})=2m_{-1}(\mathbf{q})$.
  In the long wavelength limit we have the relationship
\begin{align}
\lim_{\mathbf{q}\to0}\chi(\mathbf{q})=\rho\kappa,
\end{align}
known as the compressibility sum rule, with $\kappa$ being the isothermal compressibility\footnote{The compressibility is $\kappa=\alpha_{uu}/\rho^2(\alpha_{\rho\rho}\alpha_{uu}-\alpha_{\rho u}^2)$. Since the density-strain term is small,   $\kappa\approx1/(\rho^{2}\alpha_{\rho\rho})$.}. In Figs.~\ref{figsumrules}(c) and (d) we demonstrate that the response function calculated from the density fluctuations satisfies this relationship.

Analogously to the density response, we can similarly analyse the current fluctuations.  This leads to the static current response functions as the $p=-1$ moment of the respective current dynamical structure factors, i.e.,
\begin{align}
\chi^\sigma(\mathbf{q})=\sum_\nu\frac{2|\delta\tilde{J}^\sigma_{\nu \mathbf{q}}|^2}{N\hbar\omega_{\nu \mathbf{q}}},\qquad \sigma=\{L,T\}.
\end{align}
At all $q$ the static longitudinal current response function satisfies the equivalent of the $f$-sum rule
\begin{align}
\chi^L(q)=\frac{1}{m}.
\end{align}
The static transverse current response function allows us to identify the normal density in the
macroscopic ($q\to0$) limit  \cite{PinesNozieres1966,BECbook}
\begin{align}
\lim_{q\to0}\chi^T(q)=\frac{1}{m}\frac{\rho_n}{\rho}.\label{transcurrule}
\end{align} 
These current response function results are verified in Figs.~\ref{figsumrules}(e)-(h).

%\begin{figure}[htbp]
% \includegraphics[width=3.4in]{fig_linedata1.pdf}
%\caption{ The states analyzed are the same as in Fig.~\ref{figdenflucts}.   }\label{figlinedata}
%\end{figure}
  
 \section{Conclusions and outlook}
\label{Sec:concl}

In this work, we have developed a microscopic description of the
long-wavelength acoustic fluctuations of two-dimensional dipolar
supersolids with triangular and honeycomb crystalline order. The density,
phase, and current fluctuations are obtained directly as matrix elements of
the BdG excitation modes. A central result is a method for
extracting the lattice displacement field through a local analysis of the
stationary points of the unit-cell density profile. This displacement field
allows the total-density fluctuation to be decomposed into contributions
from crystal strain and particle transport relative to the lattice.
Together with the phase fluctuations, it also allows the current to be
resolved into its superfluid and normal components. We find that
these independently constructed quantities satisfy the two-fluid current
relation expected from zero-temperature supersolid hydrodynamics.

Despite their distinct unit-cell density profiles, the triangular and
honeycomb supersolids exhibit the same underlying organization into one
transverse and two longitudinal acoustic branches. For both crystalline
geometries, the transverse mode (\(\nu=0\)) is a lattice shear mode whose
current response directly probes the non-superfluid density. The lower
longitudinal mode (\(\nu=1\)) is a counterflow mode: its strain and
defect-density fluctuations are individually large but substantially cancel
in the total-density response, while its superfluid and normal
current contributions similarly oppose one another. By contrast, the upper
longitudinal mode (\(\nu=2\)) is a coflowing, density-dominated mode in which
the two current contributions reinforce each other. It consequently carries
most of the long-wavelength density and phase response.  

We have compared the microscopic fluctuation amplitudes with their hydrodynamic limits, which are specified entirely by the elastic coefficients of the supersolid. This extends previous microscopic tests of hydrodynamic sound speeds to the mode amplitudes and their density, phase, displacement, and current content. The calculated fluctuations also satisfy the relevant density and current sum rules and reproduce the expected static response functions.  

The predictions presented here could be explored in current experiments. Our analysis assumes a uniform planar geometry, for which the quasiparticles have well-defined quasimomentum. Comparable conditions could be realized
using box traps \cite{Navon2021a,Juhasz2022a} or toroidal geometries \cite{Sindik2024a}. In these systems, individual sound branches could be selectively excited using spatially structured optical perturbations. A
related procedure was demonstrated in Ref.~\cite{Liebster2025a}, where phase-imprinting and site-displacement perturbations were used to preferentially excite  sound waves of the different branches.
High-resolution \textit{in situ} imaging could be used to measure both the density response and the motion of the crystal sites. Density fluctuation measurements provide access to the static structure factor
\cite{Hertkorn2021a,Schmidt2021a,Blakie2023a}, while the dynamical structure
factor can be measured using Bragg spectroscopy \cite{Petter2021a,Houwman2024a}.

Measurements of current fluctuations would be particularly valuable because the transverse current couples directly to the shear branch, which is absent from the total-density response. The longitudinal current can, in principle, be inferred from time-resolved density measurements using the continuity equation. Accessing the transverse current is more challenging because it requires information about the local velocity field. Recent progress in measuring velocity correlations in quantum-gas experiments, motivated by studies of quantum turbulence \cite{Zhao2025a}, may provide a route towards
such measurements.

Several extensions of this work would be valuable. Thermal effects are
important in current experiments \cite{Yoo2010a,Hofmann2021a,Sohmen2021a,Sanchez-Baena2023a} and require a
microscopic treatment beyond the zero-temperature BdG theory used here. Possible approaches include quantum Monte Carlo \cite{Peotta2025a} and classical-field methods \cite{Linscott2014a,Bland2022a}. A detailed study of the fluctuations in the finite temperature regime could shed more light on the distinction between superfluid, normal and lattice components of the density (e.g.,~see \cite{Saslow2025a}).
Moreover, existing dipolar-supersolid experiments are generally performed in inhomogeneous harmonic traps. Applying the present analysis to these systems will require accounting for finite-size effects and the absence of exact translational symmetry.
An additional direction is provided by supersolids with tilted dipoles. Tilting the polarization away from the \(z\) axis modifies the phase diagram \cite{Lima2025a} and makes the triangular and honeycomb states anisotropic \cite{Lima2025a,Cook2026a}. The formalism developed here extends naturally to this regime, but the superfluid density and elastic coefficients become tensorial, and the acoustic modes need not have purely longitudinal or transverse character. Some related aspects of this mode mixing have already been considered for the stripe phase \cite{Poli2026a,Cook2026a}.

\section*{Acknowledgments}
PBB would like to acknowledge discussions with A-C.~Lee, D.~Baillie, and R.~Bisset. He also acknowledges use of high-performance computing resources made available through the New Zealand eScience Infrastructure (NeSI) and funding support from the Marsden Fund of the Royal Society of New Zealand. 
\noindent

 % \bibliography{dipolarpbb} 
 

\end{document}